\documentclass[aps,prl,floatfix,superscriptaddress,showpacs,twocolumn]{revtex4}
\usepackage[dvips]{graphicx}

\usepackage{longtable}
\usepackage{dcolumn}
\usepackage[dvips]{graphicx}
\usepackage{bm}
\usepackage{bbm}
\usepackage{times}
\usepackage{nicefrac}
\usepackage{amsmath}
\usepackage{amsfonts}
\usepackage{amssymb}
\usepackage{amsthm}

\newcolumntype{.}{D{x}{}{-1}}

\newcommand{\Za}{Z\alpha}

\begin{document}

\title{The $\bm g$-factor of light ions for an improved determination of the fine-structure constant}

\author{V.~A. Yerokhin}
\address{Max~Planck~Institute for Nuclear Physics, Saupfercheckweg~1, D~69117 Heidelberg, Germany}
\affiliation{Center for Advanced Studies, Peter the Great St.~Petersburg Polytechnic University,
195251 St.~Petersburg, Russia}
\author{E.~Berseneva}
\address{Max~Planck~Institute for Nuclear Physics, Saupfercheckweg~1, D~69117 Heidelberg, Germany}
\address{Department of Physics, St. Petersburg State University, 198504 St. Petersburg, Russia}
\author{Z. Harman}
\address{Max~Planck~Institute for Nuclear Physics, Saupfercheckweg~1, D~69117 Heidelberg, Germany}
\author{I.~I.~Tupitsyn}
\address{Department of Physics, St. Petersburg State University, 198504 St. Petersburg, Russia}
\author{C.~H. Keitel}
\address{Max~Planck~Institute for Nuclear Physics, Saupfercheckweg~1, D~69117 Heidelberg, Germany}

\begin{abstract}

A weighted difference of the $g$-factors of the H- and Li-like ions of the same element is
theoretically studied and optimized in order to maximize the cancelation of nuclear effects
between the two charge states. We show that this weighted difference and its combination for
two different elements can be used to extract a value for the fine-structure constant from near-future
bound-electron $g$-factor experiments with an accuracy competitive with or better than the
present literature value.

\end{abstract}


\maketitle

Precision measurements of the free-electron $g$-factor have enabled determination of the
fine-structure constant $\alpha$ to a high accuracy~\cite{hanneke:08,mohr:12:codata}. An
independent value of $\alpha$ may be extracted from the measurement of the $g$-factor of an
electron bound in an H-like ion. This can be accomplished by identifying the leading relativistic
(Dirac) contribution $ g_{\rm D} = \nicefrac{2}{3} \bigl( 1+ 2 \sqrt{1-(\Za)^2} \bigr)\,, $ with
$Z$ being the nuclear charge number, after subtracting corrections induced by quantum
electrodynamics (QED) and nuclear effects from the measured value. The sensitivity of $g_{\rm D}$
to $\alpha$ is largest for heavy ions. For these ions, however, nuclear effects (charge
distributions, polarizabilities etc.) are not well understood and set a limitation on the ultimate
accuracy of such determination.

In Ref.~\cite{shabaev:02:li}, it was suggested to use a weighted difference of the
$g$-factors of the H- and Li-like charge states of the same element in order to reduce the
nuclear size effect by about two orders of magnitude for high-$Z$ ions. In
Ref.~\cite{shabaev:06:prl} (see also \cite{hitrap:08}), a specific weighted difference of the
$g$-factors of heavy H- and B-like ions with the same $Z$ was put forward. It was demonstrated
that the theoretical uncertainty of the nuclear size effect in this difference can be brought down
to $4\times10^{-10}$ for heavy ions around Pb, which was several times smaller than the uncertainty
due to $\alpha$ at that time. Since then, however, the uncertainty of $\alpha$ was reduced by an
order of magnitude~\cite{bouchendira:11,aoyama:12,aoyama:15}, making it more difficult to
access $\alpha$ in such experiments. In this Letter we propose a weighted difference of the
$g$-factors of low-$Z$ ions, for which a much stronger cancelation of nuclear effects can be
achieved. The low-$Z$ region also seems favorable from the experimental point of view, since
experiments so far concentrated in this regime \cite{sturm:14,wagner:13,sturm:11}.

Measurements of the $g$-factor of H-like ions have reached the fractional level of
accuracy of $3\times 10^{-11}$ \cite{sturm:14}. Experiments have also been performed for Li-like
ions \cite{wagner:13}. In the future, it should be possible to perform experiments not only with a
single ion in the trap, but also with several ions simultaneously. Such a setup will directly
access differences of $g$-factors of different ions, greatly reducing systematic uncertainties and
possibly gaining two orders of magnitude in accuracy \cite{sturm:priv}. Such experiments, complemented
by corresponding improvements in the theoretical description, would become sensitive to the uncertainty
of $\alpha$.

In the present Letter we put forward a method to extract $\alpha$ to higher accuracy by employing
the weighted difference of the $g$-factors of the H- and Li-like charge states
of the same (light) element. The weight $\Xi$ of this difference will be determined by studying the
$\Za$ and $1/Z$ expansions of the finite nuclear size (fns) effects, in such a way that the
cancelation of unwanted contributions is maximized. Specifically, we introduce the following
$\Xi$-weighted difference of the $g$-factors of the Li- and H-like charge states
of the same element,
\begin{align}
\label{eq:36}
\delta_{\Xi}g = g(2s) - \Xi\, g(1s)\,,
\end{align}
where $g(2s)$ is the ground-state $g$-factor of the Li-like ion, $g(1s)$ is the ground-state $g$-factor of
the H-like ion and the weight parameter $\Xi$ is defined as
\begin{align}
\label{eq:37}
\Xi = 2^{-2\gamma-1}\,\left[ 1 + \frac{3}{16}(\Za)^2\right]  \left(1-\frac{2851}{1000}\frac1Z
 + \frac{107}{100}\frac1{Z^2}\right)\,,
\end{align}
where $\gamma  = \sqrt{1-(\Za)^2}$. The justification of such a choice of $\Xi$ will be given below
by studying the individual contributions to the fns effect.

\textit{One-electron finite nuclear size $\delta g_{\rm N}^{(0)}$.} -- The leading one-electron fns
correction is defined by the difference $\delta g_{\rm N}^{(0)} =
g_{\rm ext}^{(0)} - g_{\rm pnt}^{(0)}\,, $ where $g_{\rm ext}^{(0)}$ and $g_{\rm pnt}^{(0)}$ are
the relativistic one-electron $g$-factors evaluated with the extended and the point nuclear models,
respectively. For $ns$ states, they are given by the radial integral
\begin{equation} \label{eq:03}
g^{(0)} = -\frac83\, \int_0^{\infty}dr\,r^3\,g_a(r)\,f_a(r)\,,
\end{equation}
where $g_a$ and $f_a$ are the upper and the lower radial components of the reference-state wave
function, respectively.

We parameterize the leading one-electron fns effect for $ns$ states as
\begin{align} \label{eq:04}
\delta g_{\rm N}^{(0)}(ns) = &\ \frac{2}{5}\,\left( \frac{2\,\Za\, R_{\rm sph}}{n}\right)^{2\gamma}\,
  \frac{(\Za)^2}{n}\,
  \left[ 1 + (\Za)^2\,H_n^{(0,2+)}\right]\,,
\end{align}
where $R_{\rm sph} = \sqrt{5/3}\,R$ is the radius of the sphere with the root-mean-square (rms)
radius $R$ and $H_n^{(0,2+)}$ is the remainder induced by relativistic effects. The superscript
$(0,2+)$ indicates the contribution of zeroth order in $1/Z$ and second and higher orders in $\Za$.
The nonrelativistic limit of Eq.~(\ref{eq:04}) agrees with the known result
\cite{karshenboim:00:pla}. The leading relativistic correction $H_n^{(0,2)}$ was derived
analytically in Ref.~\cite{glazov:01:pla}. From that work, we deduce that the difference of the
leading relativistic corrections for the $2s$ and $1s$ states is just a constant,
\begin{align} \label{eq:05}
H_{21}^{(0,2)} \equiv H_2^{(0,2)} - H_1^{(0,2)} = \frac{3}{16} \,.
\end{align}

In the present work we calculated $\delta g_{\rm N}^{(0)}$ numerically. The Dirac equation for the
extended nucleus was solved with the Dual Kinetic Balance (DKB) method \cite{shabaev:04:DKB}. In
order to compensate large numerical cancelations occurring in the low-$Z$ region, we
implemented the DKB method in quadruple-precision (32-digit) arithmetics. After that, we were able to
determine $\delta g_{\rm N}^{(0)}$ and, therefore, $H_n^{(0,2+)}$ to a very high accuracy.

We found that the model dependence of the relativistic fns correction $H_n^{(0,2+)}$ is generally
not small; it varies from 1\% in the medium-$Z$ region to 5\% in the low-$Z$ region. On the
contrary, the model dependence of the difference $H_2^{(0,2+)}-H_1^{(0,2+)}$ is very weak. We thus
obtain that both the model dependence and the $R$ dependence of $\delta g_{\rm N}^{(0)}(ns)$ can be
canceled up to a very high accuracy by forming a suitably chosen difference. Specifically, the
following difference of the one-electron $g$-factors cancels the one-electron fns contributions of
relative orders $(\Za)^0$ and $(\Za)^2$,
\begin{align} \label{eq:06}
\delta_{\Xi_0}g = g^{(0)}(2s) - \Xi_0\, g^{(0)}(1s)\,,
\end{align}
with
\begin{align} \label{eq:07}
\Xi_0 = 2^{-2\gamma-1}\,\left[ 1 + \frac{3}{16}(\Za)^2\right]\,.
\end{align}

\textit{One-electron QED fns correction $\delta g_{\rm NQED}^{(0)}$.} -- The one-electron QED fns
correction, arising from the one-loop self-energy and vacuum polarization diagrams, can be
conveniently parameterized in terms of the dimensionless function $G_{\rm NQED}^{(0)}$
\cite{yerokhin:13:jpb},
\begin{equation} \label{eq:qed1}
\delta g_{\rm NQED}^{(0)} = \delta g_{\rm N}^{(0)}\, \frac{\alpha}{\pi}\, G^{(0)}_{\rm NQED}(\Za,R)\,,
\end{equation}
where $\delta g_{\rm N}^{(0)}$ is the leading-order fns correction. The QED fns correction was
studied in detail in \cite{yerokhin:13:jpb}, where we reported
results for the $1s$ state of H-like ions. In the present work, we extend those calculations to the
$2s$ state. The numerical results obtained are listed in Table~\ref{tab:qed_fns}. We observe that
the QED fns corrections for the $1s$ and $2s$ states, expressed in terms of the function $G_{\rm
NQED}^{(0)}$, are very close to each other. Therefore, they are significantly canceled in the
difference $\delta_{\Xi_0}g$ introduced by Eq.~(\ref{eq:06}).

\begin{table}
\caption{One-electron QED fns corrections to the bound-electron $g$-factor, expressed in terms of
 $G_{\rm NQED}^{(0)}$ defined by Eq.~(\ref{eq:qed1}). Abbreviations are as follows: "NSE" denotes
self-energy fns contribution for the $2s$ state,
"NVP" denotes the vacuum-polarization fns correction for the $2s$ state,
"2s" is the total QED fns correction for the $2s$ state, "1s" is the total
QED fns correction for the $1s$ state.
 \label{tab:qed_fns} }
\begin{center}
\begin{ruledtabular}
\begin{tabular}{l.....}
                $Z$
                & \multicolumn{1}{c}{NSE}
                & \multicolumn{1}{c}{NVP}
                & \multicolumn{1}{c}{2s}
                & \multicolumn{1}{c}{1s} \\
\hline\\[-5pt]
  6 & -0.x54\,(20)   & 0.1x58\,(1)    & -0.3x8\,(20)   & -0.x60\,(1) \\
  8 & -0.x77\,(10)   & 0.2x26\,(1)    & -0.5x5\,(10)   & -0.x70\,(1)   \\
 10 & -0.x94\,(4)    & 0.2x96\,(1)    & -0.6x5\,(4)    & -0.x807\,(9)  \\
 12 & -1.x14\,(4)    & 0.3x73\,(2)    & -0.7x7\,(4)    & -0.x905\,(8)  \\
 14 & -1.x32\,(4)    & 0.4x59\,(2)    & -0.8x6\,(4)    & -0.x996\,(5)  \\
 20 & -1.x86\,(4)    & 0.7x40\,(4)    & -1.1x2\,(4)    & -1.x237\,(3)  \\
 25 & -2.x36\,(4)    & 1.0x12\,(4)    & -1.3x5\,(4)    & -1.x404\,(2)  \\
 30 & -2.x82\,(4)    & 1.3x18\,(6)    & -1.5x0\,(4)    & -1.x542\,(2)  \\
 35 & -3.x27\,(2)    & 1.6x54\,(8)    & -1.6x2\,(4)    & -1.x655\,(1)  \\
 40 & -3.x75\,(2)    & 2.0x37\,(8)    & -1.7x1\,(2)    & -1.x733\,(1)  \\
 45 & -4.x23\,(1)    & 2.4x45\,(5)    & -1.7x9\,(2)    & -1.x793\,(1)       \\
 50 & -4.x73\,(1)    & 2.9x00\,(7)   & -1.8x3\,(1)    & -1.x821\,(1)       \\
 55 & -5.x25\,(1)    & 3.4x00\,(8)   & -1.8x5\,(1)    & -1.x819\,(1)  \\
 60 & -5.x79\,(2)    & 3.9x58\,(9)   & -1.8x3\,(2)    & -1.x780\,(1)       \\
\end{tabular}
\end{ruledtabular}
\end{center}
\end{table}

\textit{One-photon exchange fns correction $\delta g^{(1)}_N$.} -- The one-photon exchange fns
correction is the leading two-electron contribution to the fns effect. It is suppressed by the
factor of $1/Z$ with respect to the leading fns contribution $\delta g_{\rm N}^{(0)}$. The first
calculation of the one-photon exchange correction was demonstrated in Ref.~\cite{shabaev:02:li}.
Here we redid these calculations with enhancing the numerical accuracy by several orders of
magnitude, which was necessary for an accurate identification of the fns effect in the low-$Z$
region.

The one-photon exchange fns correction $\delta g^{(1)}_N$ is conveniently parameterized as follows:
\begin{align} \label{eq:26}
 \delta g^{(1)}_N = \delta g^{(0)}_N\, \frac1Z \,\bigl[ H^{(1,0)} + (\Za)^2 \,H^{(1,2+)}\bigr]\,,
\end{align}
where $\delta g^{(0)}_N$ is the one-electron fns correction given by Eq.~(\ref{eq:04}), $H^{(1,0)}$
is the leading nonrelativistic contribution and $H^{(1,2+)}$ is the higher-order remainder.

Results of our numerical calculations of $\delta g^{(1)}_N$ for different $Z$ values were
extrapolated to $Z\to 0$, yielding an accurate value of the nonrelativistic contribution, $
H^{(1,0)}= -2.8512\,(10)\,.$ Our calculations confirm that $H^{(1,0)}$ depends neither on the
nuclear model nor on the charge radius. Because of this, the nonrelativistic contribution can be
exactly canceled in the weighted difference. Specifically, we conclude that the following
difference of the $g$-factors cancels the dominant part of the $1/Z$ fns contribution for light
ions,
\begin{align} \label{eq:28a}
\delta_{\Xi_1}g = g^{(1)}(2s) - \Xi_0\, \left(-\frac{2851}{1000}\frac1Z\right) g^{(0)}(1s)\,.
\end{align}

\textit{Two- and more photon exchange fns correction $\delta g^{(2+)}_N$.} -- The fns correction
with two and more photon exchanges is suppressed by the factor of $1/Z^2$ as compared to the
one-electron fns contribution. It can be parameterized analogously to Eq.~(\ref{eq:26}),
\begin{align} \label{eq:312}
 \delta g^{(2+)}_N = \delta g^{(0)}_N\, \frac1{Z^2} \,\bigl[ H^{(2+,0)} + (\Za)^2 \,H^{(2+,2+)}\bigr]\,.
\end{align}
In order to access $\delta g^{(2+)}_N$ in a numerical calculation, one needs to calculate the two-
and more photon exchange correction for the extended and the point nuclear models and take the
difference, $
 \delta g^{(2+)}_N = \delta g^{(2+)}_{\rm ext} - \delta g^{(2+)}_{\rm pnt}\,.
$ Here we performed these calculations within the Breit approximation, in three steps. Firstly, we
solved the no-pair Dirac-Coulomb-Breit Hamiltonian by the Configuration-Interaction
Dirac-Fock-Sturm method \cite{tupitsyn:03}. Secondly, we subtracted the leading-order
contributions of orders $1/Z^0$ and $1/Z^1$ (calculated separately by perturbation theory), thus
identifying the contribution of order $1/Z^2$ and higher. Thirdly, we repeated the calculation for
the extended and the point nuclear models and, by taking the difference, obtained
$\delta g^{(2+)}_N$.

By performing numerical calculations for a series of ions and fitting the numerical results, we
obtained the nonrelativistic value of the $1/Z^2$ correction as $ H^{(2,0)}= 1.070\,(25)\,.$ We
also found that $H^{(2,0)}$ does not depend on the nuclear radius and, therefore, can be canceled
in the difference. We conclude that the following difference cancels the
dominant part of the $1/Z^2$ fns contribution for low-$Z$ ions:
\begin{align} \label{eq:28b}
\delta_{\Xi_2}g = g^{(2)}(2s) - \Xi_0\, \left(\frac{107}{100}\frac1{Z^2}\right) g^{(0)}(1s)\,.
\end{align}

\begin{figure*}[t]
\centerline{
\resizebox{\textwidth}{!}{%
  \includegraphics{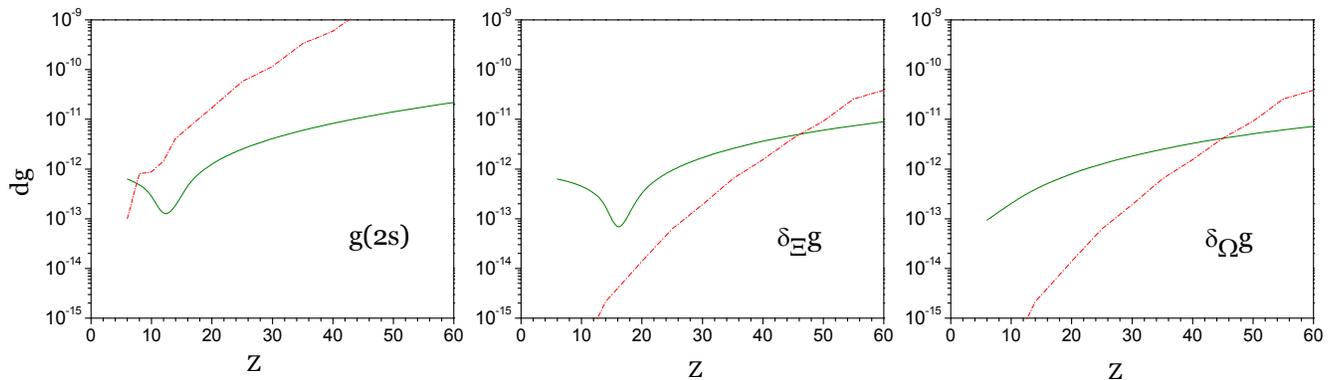}
}}
 \caption{(Color online) Comparison of the error $dg=\frac{\partial g}{\partial \alpha} \delta\alpha$ due to the uncertainty
 of the fine-structure constant $\delta \alpha/\alpha =
3.2\times 10^{-10}$ (solid line, green) and the error due to the
 finite nuclear size effect (dashed-dot line, red), for the
 $g$-factor of the ground state of Li-like ions $g(2s)$ (left graph); for the weighted difference $\delta_{\Xi}g(Z)$ (middle graph);
 and for the weighted difference $\delta_{\Omega}g = \delta_{\Xi}g(Z) - \delta_{\Xi}g([Z/2])$ (right graph).
\label{fig:sens}}
\end{figure*}

\textit{The weighted difference of the $2s$ and $1s$ $g$-factors.} -- Combining the weighted
differences for specific fns contributions, Eqs.~(\ref{eq:06}), (\ref{eq:28a}), and (\ref{eq:28b}),
we arrive at the total $\Xi$-weighted difference of the $g$-factors of the Li- and H-like charge
states in Eq.~(\ref{eq:36}), with the weight parameter $\Xi$ defined by Eq.~(\ref{eq:37}). Basing
on the preceding analysis, we claim that the $\Xi$-weighted difference $\delta_{\Xi}g$ cancels the
nonrelativistic fns corrections to order $1/Z^0$, $1/Z^1$, and $1/Z^2$ and, in addition, the
nuclear model-dependent contribution to order $(Z\alpha)^2/Z^0$. A small remaining fns correction
to $\delta_{\Xi}g$ is calculated numerically. The leading nuclear model dependence of this
correction comes only in the relative order $(Z\alpha)^2/Z$, with a numerically small coefficient.

We now address the question whether the weighted difference $\delta_{\Xi}g$ in light ions might
be useful for the determination of $\alpha$. Expanding $\delta_{\Xi}g$ in $\alpha$ and keeping
$\Xi$ fixed, we have
\begin{align}\label{eq42}
\delta_{\Xi}g =
2\,(1-\Xi) -\frac23(\Za)^2 \left(\frac14-\Xi\right) + \frac{\alpha}{\pi}(1-\Xi) + \ldots\,.
\end{align}
The sensitivity of $\delta_{\Xi}g$ on $\alpha$ comes, first, from the binding effects (the second
term in the right-hand-side) and, second, from the free-electron QED effects (the third term). By
varying $\alpha$ in Eq.~(\ref{eq42}) within its current error bars $\delta \alpha/\alpha =
3.2\times 10^{-10}$ \cite{mohr:12:codata}, we obtain the corresponding error of $\delta_{\Xi}g$.

In Fig.~\ref{fig:sens} we compare the uncertainty due to $\alpha$ to the nuclear-model and
nuclear-radius error of the fns effect, keeping in mind that the latter sets the ultimate
limit of the accuracy of theoretical calculations. The left panel of Fig.~\ref{fig:sens} represents
this comparison for the $g$-factor of Li-like ions $g(2s)$, whereas the middle panel
gives the same comparison for the $\Xi$-weighted difference $\delta_{\Xi}g$. The dip of the
$\alpha$-sensitivity around $Z = 16$ is due to the fact that the dependences of the binding and the
free-QED effects on $\alpha$ in Eq.~(\ref{eq42}) have different signs and cancel each other.

From Fig.~\ref{fig:sens} we observe that the cancelation of the fns effects in the $\Xi$-weighted
difference $\delta_{\Xi}g$ yields an improvement of the ultimate limit of the achievable accuracy
by about 3 orders of magnitude as compared to the $g$-factor $g(2s)$ of the Li-like ion. Up to $Z
\approx 45$, the weighted difference $\delta_{\Xi}g$ yields possibilities for an improved
determination of $\alpha$.

The extraction of $\alpha$ from $\delta_{\Xi}g$ may be argued to have two disadvantages. The
first is the unfortunate cancelation of the $\alpha$-dependence around $Z = 16$, which leads to a
loss of sensitivity to $\alpha$ of up to an order of magnitude. The second is that the theory of
$\delta_{\Xi}g$ contains the same free-QED part as the determination of $\alpha$ from the
free-electron $g$-factor, meaning that these two methods are not completely independent.
Both disadvantages can be circumvented by forming another difference,
\begin{align}
\label{eq:Omega}
\delta_{\Omega}g = \delta_{\Xi}g(Z) - \delta_{\Xi}g([Z/2])\,,
\end{align}
where $\delta_{\Xi}g(Z)$ is the $\Xi$-weighted difference (\ref{eq:36}) for the atomic number $Z$,
and $\delta_{\Xi}g([Z/2])$ is the corresponding difference for the atomic number $[Z/2]$
($[\ldots]$ means the upper or lower integer part). By forming the difference
$\delta_{\Omega}g$, we cancel the free-QED contributions (which do not depend on $Z$), thus leaving
the leading dependence on $\alpha$ through the binding effects only and rendering the determination of
$\alpha$ independent from that employing the free-electron $g$ factor.

The comparison of the uncertainty due to $\alpha$ and the error of the fns effect for the weighted
difference $\delta_{\Omega}g$ is presented in the right panel of Fig~\ref{fig:sens}. As expected,
we find a smooth dependence of the sensitivity to $\alpha$ on $Z$, without any structure around $Z
= 16$. We observe that in the region of nuclear charges $Z = 10 - 20$, the weighted difference
$\delta_{\Omega}g$ offers better possibilities for determining $\alpha$ than $\delta_{\Xi}g$.
$\delta_{\Omega}g$ can be effectively determined in an experiment by measuring two differences,
$g(1s,Z)-g(1s,[Z/2])$ and $g(2s,Z)-g(2s,[Z/2])$, and $g(1s,[Z/2])$, where the latter is suppressed
in the weighted difference by a factor of $\Xi(Z)-\Xi([Z/2]) \approx 0.02$-$0.04$ in the region
of interest. Thus, the experimental error of $\delta_{\Omega}g$ can be improved by more than an
order of magnitude as compared to absolute $g$-factors.

We now turn to the experimental consequences of our calculations. The only element for which the
weighted difference $\delta_{\Xi}g$ has been measured is Si ($Z$=14). In Table~\ref{tab:si} we
collect all available theoretical contributions to $\delta_{\Xi}g(^{28}\mbox{\rm Si})$, including
our present results for the fns effect. Corrections due to the one- and two-electron QED
effects were taken from
Refs.~\cite{yerokhin:04,glazov:04:pra,pachucki:05:gfact,volotka:09,glazov:10,aoyama:15,volotka:14}.
Theoretical results are compared with the experimental value~\cite{sturm:13:Si,wagner:13,sturm:11}.
The error of the Dirac value and the free-QED correction is due to the uncertainty of the current
value of $ \alpha^{-1} = 137.035\,999\,074\,(44)\,$ \cite{mohr:12:codata}. The indicated
uncertainty of the fns effect of $4\times 10^{-13}$ is already smaller than the uncertainty of the
Dirac value due to $\alpha$; it is of purely numerical origin and thus can be improved further in
future calculations.

Table~\ref{tab:si} shows that the present experimental and theoretical precision of
$\delta_{\Xi}g({\rm Si})$ is on the level of few parts in $10^{-9}$, which is much lower than the
precision achieved for other systems (in particular, in H-like C the present experimental and
theoretical uncertainties are, correspondingly, $6\times 10^{-11}$ and $6\times 10^{-12}$
\cite{sturm:14}). Such underperformance, however, was more due to a lack of motivation than due to
principal obstacles. On the experimental side, the same precision as for H-like C can be
obtained for Li-like C [and, therefore, $\delta_{\Xi}g({\rm C})$], with the existing apparatus
\cite{sturm:priv}. Moreover, further experimental advance is anticipated  that could bring one or
two orders of magnitude of improvement \cite{sturm:priv}. On the theoretical side, the modern
nonrelativistic quantum electrodynamics (NRQED) approach (see, e.g., \cite{puchalski:14}) can
provide a theoretical prediction for Li-like C with the same accuracy as obtained for H-like
C \cite{pachucki:priv}. Moreover, further theoretical advance is possible, namely: the
two-loop QED correction of order $\alpha^2(\Za)^5$ and the three-loop QED correction of order
$\alpha^3(\Za)^4$ can be calculated, both for H-like and Li-like ions \cite{pachucki:priv}.

Since we are presently interested in light ions, the best way for the theoretical advance is a
combination of two complementary methods. The first one is NRQED (used, e.g., in
\cite{yan:01:prl}) that accounts for the nonrelativistic electron-electron interactions to all
orders but expands the QED and relativistic effects in $\alpha$ and $\Za$. The second method (used,
e.g., in \cite{glazov:04:pra,volotka:09,glazov:10,volotka:14}) accounts for the relativistic
effects to all orders in $\Za$ but uses perturbation expansions for QED effects (in $\alpha$) and
for the electron-electron interaction (in $1/Z$). Matching the coefficients of $\Za$ and $1/Z$
expansions of these two methods allows one to merge them together, as it was done for the energy
levels in \cite{yerokhin:10}. As a result, only higher-order corrections in $\Za$ will be expanded
in $1/Z$ and higher-order corrections in $1/Z$ will be expanded in $\Za$, whereas the lower-order
terms will be obtained complete both in $\Za$ and in $1/Z$. Such approach should allow one to
advance theory to the level required for the improved determination of $\alpha$.

It is important that the theoretical progress in calculations of $\delta_{\Xi}g$ is not hampered
by the nuclear size, and in fact by any other nuclear effects. Recent investigations
of the nuclear deformation and polarization effects \cite{zatorsky:12,volotka:14} demonstrated that
these effects are of the same short-range nature  as the nuclear size effects and that they are
canceled in the same weighted difference. In particular, we estimate the magnitude of the nuclear
polarization contribution to $\delta_{\Xi}g({\rm C})$ to be $\sim$$5\times 10^{-16}$, which is
completely negligible at the level of interest.

\begin{table}
\caption{Individual contributions to the weighted difference $\delta_{\Xi}g$ for $^{28}$Si,
$M/m = 50984.8327(3)$, $\Xi = 0.101136233077060$. \label{tab:si}}
\begin{center}
\begin{tabular}{ll.}
\hline
\hline\\[-0.25cm]
                \multicolumn{1}{l}{Contribution}
                & \multicolumn{1}{l}{Order}
                & \multicolumn{1}{c}{Value}
 \\
\hline\\[-9pt]
    Dirac        &&      1.796\,687\,x854\,216\,5\,(7) \\[2pt]
  $1$-loop QED
 &   $\alpha(\Za)^0$     &     0.002\,087\,x898\,255\,0\,(7)   \\
&    $\alpha(\Za)^2$     &     0.000\,000\,x601\,506\,0     \\
&    $\alpha(\Za)^4$     &     0.000\,000\,x014\,797\,0     \\
&    $\alpha(\Za)^{5+}$  &     0.000\,000\,x015\,48\,(52)   \\[2pt]
  $2$-loop QED
  &  $\alpha^2(\Za)^0$     &    -0.000\,003\,x186\,116\,6    \\
  &  $\alpha^2(\Za)^2$     &    -0.000\,000\,x000\,917\,9    \\
  &  $\alpha^2(\Za)^4$     &    -0.000\,000\,x000\,084\,4    \\
  &  $\alpha^2(\Za)^{5+}$  &     0.000\,000\,x000\,00\,(13)    \\[2pt]
  $\geq 3$-loop QED
  & $\alpha^3(\Za)^0$      &     0.000\,000\,x026\,517\,7\,(1)   \\
  &  $\alpha^3(\Za)^2$     &     0.000\,000\,x000\,007\,6      \\
  &  $\alpha^3(\Za)^{4+}$  &     0.000\,000\,x000\,000\,0\,(11)   \\[2pt]
   Recoil:                 &
                           &     0.000\,000\,x030\,5\,(10)     \\[2pt]
    1-photon exchange&
                            &  0.000\,321\,x590\,803\,3   \\
    2-photon exchange&
                            & -0.000\,006\,x876\,0\,(5)     \\
$\geq 3$-photon exchange &
                               &  0.000\,000\,x093\,0\,(60)    \\[2pt]
    2-electron QED   &
                          & -0.000\,000\,x236\,0\,(50)     \\
    2-electron recoil&
                         & -0.000\,000\,x012\,1\,(7)      \\[2pt]
Finite nuclear size &&  -0.000\,000\,x000\,000\,6\,(4)  \\[2pt]
\hline\\[-9pt]
    Total theory    &&   1.799\,087\,x813\,9\,(79)     \\
    Experiment  \cite{sturm:13:Si,wagner:13,sturm:11}
                    &&   1.799\,087\,x812\,5\,(21)   \\
\hline
\hline\\[-0.5cm]
\end{tabular}
\end{center}
\end{table}

In summary, specific weighted differences of the $g$-factors of H- and Li-like ions were
investigated. The weight parameter $\Xi$ was determined by the condition of cancelation of the
nonrelativistic finite nuclear size corrections to order $1/Z^0$, $1/Z^1$, and $1/Z^2$ and, in
addition, the relativistic nuclear model-dependent contribution to order $(Z\alpha)^2/Z^0$. We
demonstrated that the $\Xi$- and $\Omega$-weighted differences~(\ref{eq:36}) and (\ref{eq:Omega})
can be used for an effective cancelation of nuclear structural effects. This independent scheme may
be used to extract the fine-structure constant from a comparison of experimental and theoretical
bound-electron $g$-factors with an accuracy competitive with or better than the current value.

\begin{acknowledgements}
We acknowledge insightful conversations with Sven Sturm.
V.A.Y.  acknowledges  support by Ministry of Education and Science of Russian Federation (program
for organizing and carrying out scientific investigations). E.B. acknowledges support from the
German-Russian Interdisciplinary Science Center, project No. P-2014a-9.
\end{acknowledgements}

\end{document}